# A new perspective on Einstein's philosophy of cosmology

Cormac O'Raifeartaigh

*School of Science and Computing, Waterford Institute of Technology, Cork Road, Waterford, Ireland*

Correspondence: coraifeartaigh@wit.ie

## 1. Introduction

It has recently been discovered that Einstein once attempted – and subsequently abandoned – a 'steady-state' model of the expanding universe (O'Raifeartaigh *et al* 2014; O'Raifeartaigh 2014; Nussbaumer 2014a). An unpublished manuscript on the Albert Einstein Online Archive (Einstein 1931a) demonstrates that Einstein explored the possibility of a universe that expands but remains essentially unchanged due to a continuous formation of matter from empty space (figure 1). Several aspects of the manuscript indicate that it was written in the early months of 1931, during Einstein's first trip to California, and the work therefore probably represents Einstein's first attempt at a theoretical model of the cosmos in the wake of emerging evidence for an expanding universe (O'Raifeartaigh *et al* 2014; Nussbaumer 2014a). It appears that Einstein abandoned the idea when he discovered that his steady-state model led to a null solution, as described below.

Many years later, steady-state models of the expanding cosmos were independently proposed by Fred Hoyle, Hermann Bondi and Thomas Gold (Hoyle 1948; Bondi and Gold 1948). The hypothesis formed a well-known alternative to 'big bang' cosmology for many years (North 1965 pp 208-222; Kragh 1996 pp 186-218; Nussbaumer and Bieri 2009 pp 161-163), although it was eventually ruled out by observations such as the distribution of the galaxies at different epochs and the cosmic microwave background (Kragh 1996 pp 318-380; Kragh 2007 pp 201-206; Narlikar 1988 p219). While it could be argued that steady-state cosmologies are of little practical interest today, we find it most interesting that Einstein conducted an internal debate between steady-state and evolving models of the cosmos decades before a similar debate engulfed the cosmological community. In particular, the episode offers several new insights into Einstein's cosmology, from his view of the role of the cosmological constant to his attitude to the question of cosmic origins. More generally, Einstein's exploration of steady-state cosmology casts new light on his philosophical journey from a static, bounded cosmology to the dynamic, evolving universe, and is indicative of a pragmatic, empiricist approach to cosmology.

## 2. Historical context

Following the successful formulation of his general theory of relativity (Einstein 1915, 1916), Einstein lost little time in applying his new theory of gravity, space and time to the universe as a whole. A major motivation was the clarification of the conceptual foundations of general relativity, i.e., to establish "whether the relativity concept can be followed through to the finish, or whether it leads to contradictions" (Einstein 1917a). Assuming a cosmos that was homogeneous, isotropic and static over time,[1] and that a consistent theory of gravitation should incorporate Mach's principle,[2] he found it necessary to add a new 'cosmological constant' term to the field equations of relativity in order to predict a universe with a non-zero density of matter (Einstein 1917b). This approach led Einstein to a finite, static cosmos of spherical spatial geometry whose radius was directly related to the density of matter.

That same year, the Dutch theorist Willem de Sitter noted that general relativity allowed another model of the cosmos, namely the case of a universe empty of matter (de Sitter 1917). Einstein was greatly perturbed by de Sitter's solution as it suggested a spacetime metric that was independent of the matter it contained, in conflict with his understanding of Mach's principle (Einstein 1918b). The de Sitter model became a source of some confusion amongst theorists for some years; it was later realised that the model was not static (Weyl 1923; Lemaître 1925). However, the solution attracted some attention in the 1920s because it predicted that the radiation emitted by objects inserted as test particles into the 'empty' universe would be red-shifted, a prediction that chimed with emerging astronomical observations of the spiral nebulae.[3]

In 1922, the young Russian physicist Alexander Friedman suggested that non-stationary solutions to the Einstein field equations should be considered in relativistic models of the cosmos (Friedman 1922). With a second paper in 1924, Friedman explored almost all the main theoretical possibilities for the evolution of the cosmos and its geometry (Friedman 1924). However, Einstein did not welcome Friedman's time-varying models of the cosmos. His first reaction was that Friedman had made a mathematical error (Einstein 1922). When Friedman showed that the error lay in Einstein's correction, Einstein duly retracted it (Einstein 1923a); however, an unpublished draft of Einstein's retraction makes it clear that he considered time-varying models of the cosmos unrealistic (Einstein 1923b: Stachel 1977: Nussbaumer and Bieri 2009 pp 91-92).

[1] No empirical evidence for a non-static universe was known to Einstein at the time.

[2] Einstein's view of Mach's principle in these years was that space could not have an existence independent of matter and thus the spatial components of the metric tensor should vanish at infinity (Einstein 1918a; Janssen 2005)

[3] Observations of the redshifts of the spiral nebulae were published by VM Slipher in 1915 and 1917 (Slipher 1915, 1917), and became widely known when they were included in a well-known book on relativity (Eddington 1923).

Unaware of Friedman's analysis, the Belgian physicist Georges Lemaître proposed an expanding model of the cosmos in 1927. A theoretician with significant training in astronomy, Lemaître was aware of V.M. Slipher's observations of the redshifts of the spiral nebulae (Slipher 1915, 1917) and of Edwin Hubble's emerging measurements (Hubble 1925) of the vast distances to the nebulae (Kragh 1996 p29; Farrell 2005 p90). Interpreting Slipher's redshifts as a relativistic expansion of space, Lemaître showed that a universe of expanding radius could be derived from Einstein's field equations, and estimated a rate of cosmic expansion from average values of the velocities and distances of the nebulae from Slipher and Hubble respectively. This work received very little attention at first, probably because it was published in French in a little-known Belgian journal (Lemaître 1927). However, Lemaître discussed the model directly with Einstein at the 1927 Solvay conference, only to have it dismissed with the forthright comment:"Vos calculs sont corrects, mais vôtre physique est abominable" (Lemaître 1958).

In 1929, Edwin Hubble published the first empirical evidence of a linear relation between the redshifts of the spiral nebulae (now known to be extra-galactic) and their radial distance (Hubble 1929). By this stage, it had also been established that the static models of Einstein and de Sitter presented problems of a theoretical nature: Einstein's universe was not stable against perturbation (Lemaître 1927; Eddington 1930) while de Sitter's universe was not static (Weyl 1923; Lemaître 1925). In consequence, theorists began to take Lemaître's model seriously, and a variety of time-varying relativistic models of the cosmos of the Friedman-Lemaître type were advanced (Eddington 1930, 1931: de Sitter 1930a, 1930b; Tolman 1930a, 1930b, 1931, 1932; Heckmann 1931, 1932; Robertson 1932, 1933).

By 1931, Einstein had accepted the dynamic universe. During a three-month sojourn at Caltech in Pasadena in early 1931, a trip that included discussions with the astronomers of Mount Wilson Observatory and with the Caltech theorist Richard Tolman,[4] Einstein made several public statements to the effect that he viewed Hubble's observations as likely evidence for a cosmic expansion. For example, the *New York Times* reported Einstein as commenting that "New observations by Hubble and Humason concerning the redshift of light in distant nebulae makes the presumptions near that the general structure of the universe is not static" (AP 1931). Not long afterwards, Einstein published two distinct dynamic models of the cosmos, the Friedman-Einstein model of 1931 and the Einstein-de Sitter model of 1932 (Einstein 1931b; Einstein and de Sitter 1932).

[4] An account of Einstein's time in Pasadena can be found in (Nussbaumer and Bieri 2009) pp 144-146 and (Bartusiak 2009) pp 251-256.

Written in April 1931, the Friedman-Einstein model marked the first scientific publication in which Einstein formally abandoned the static universe. Citing Hubble's observations, he suggested that the assumption of a static universe was no longer justified (Einstein 1931b). Adopting Friedman's 1922 analysis of a universe of time-varying radius and positive spatial curvature, Einstein then removed the cosmological constant on the grounds that it was both unsatisfactory (it gave an unstable solution) and unnecessary. The resulting model predicted a cosmos that would undergo an expansion followed by a contraction, and Einstein made use of Hubble's observations to extract estimates for the current radius of the universe, the mean density of matter and the timespan of the expansion (Einstein 1931b). Noting that the latter estimate was less than the ages of the stars estimated from astrophysics, Einstein attributed the problem to errors introduced by the simplifying assumptions of the model, notably the assumption of homogeneity.[5]

In early 1932, Einstein and Willem de Sitter proposed an alternative model of the expanding universe, based on Otto Heckmann's observation that a finite density of matter in a non-static universe does not necessarily demand a curvature of space (Heckmann 1931). Mindful of a lack of empirical evidence for spatial curvature, Einstein and de Sitter set this parameter to zero (Einstein and de Sitter 1932). With both the cosmological constant and spatial curvature removed, the resulting model described a cosmos of Euclidean geometry in which the rate of expansion $h$ was related to the mean density of matter $\rho$ by the simple relation $h^2 = \frac{1}{3}\kappa\rho$, where $\kappa$ is the Einstein constant. Applying Hubble's value of 500 km $s^{-1}$ $Mpc^{-1}$ for the recession rate of the galaxies, the authors calculated a value of $4 \times 10^{-28}$ $gcm^{-3}$ for the mean density of matter, a value that they found reasonably compatible with estimates from astronomy (Einstein and de Sitter 1932).

The Einstein-de Sitter model became very well-known and it played a significant role in the development of 20th century cosmology (North 1965 p134; Kragh 1996 p35; Nussbaumer and Bieri 2009 p152; Nussbaumer 2014b). One reason was that it marked an important hypothetical case in which the expansion of the universe was precisely balanced by a critical density of matter; a cosmos of lower mass density would be of hyperbolic geometry and expand at an ever-increasing rate, while a cosmos of higher mass density would be of spherical geometry and eventually collapse. Another reason was the model's great simplicity; in the absence of any empirical evidence for spatial curvature or a cosmological constant, there was little reason to turn to more complicated models. However, the timespan of the expansion was not considered in the rather terse paper. We recently discovered a little-known paper by Einstein containing a review of the Einstein-de Sitter model (Einstein 1933a; O'Raifeartaigh et al. 2015): as in the case of the Friedman-Einstein model, it is noted that the time of expansion is less than the estimated ages of the stars and the problem is attributed to the simplifying assumptions of the model.

[5] We have recently presented an analysis and first English translation of this work. We find that all of Einstein's estimates contain a systematic numerical error (O'Raifeartaigh and McCann 2014).

## 3. Einstein's steady-state manuscript

As pointed out in the introduction, it appears that Einstein's steady-state manuscript was written in early 1931, before the Friedman-Einstein model of April 1931. The manuscript (Einstein 1931a) opens with a brief discussion of what Einstein terms the 'cosmological problem', i.e., the problem of gravitational collapse in classical and relativistic models of the universe: "It is well known that the most important fundamental difficulty that emerges when one asks how the stellar matter fills up space in very large dimensions is that the laws of gravity are not in general consistent with the hypothesis of a finite mean density of matter. Thus, at a time when Newton's theory of gravity was still generally accepted, Seeliger had already modified the Newtonian law by the introduction of a distance function that, for large distances r, diminishes considerably faster than $1/r^2$." Noting a similar problem in general relativity, Einstein recalls his introduction of the cosmological constant to the field equations in order to allow the prediction of a universe of constant radius and non-zero density of matter: "This difficulty also arises in the general theory of relativity. However, I have shown in the past that this can be overcome by the introduction of the so-called "λ–term" to the field equations. The field equations can then be written in the form

$$\left(R_{ik} - \frac{1}{2} g_{ik} R\right) - \lambda g_{ik} = \kappa T_{ik} \qquad \ldots.(1)$$

…At that time, I showed that these equations can be satisfied by a spherical space of constant radius over time, in which matter has a density ρ that is constant over space and time."

Einstein then notes that his static model was invalidated on both theoretical and observational grounds. In the first instance, the static model was unstable, while dynamic solutions existed: "On the one hand, it follows from investigations based on the same equations by [ ] and by Tolman [6] that there also exist spherical solutions with a world radius P that is variable over time, and that my solution is not stable with respect to variations of P over time." Second, the astronomical observations of Edwin Hubble changed the playing field: "On the other hand, Hubbel's [sic] exceedingly important investigations have shown that the extragalactic nebulae have the following two properties:1)Within the bounds of observational accuracy they are uniformly distributed in space 2)They possess a Doppler effect proportional to their distance."

Einstein then points out that the time-varying solutions of the field equations proposed by de Sitter and Tolman are consistent with Hubble's observations, but predict a timespan for the expansion that is problematic: "De Sitter and Tolman have already shown that there are solutions to equations (1) that can account for these observations. However the difficulty arose that the theory unvaryingly led to a beginning in time about $10^{10}$-$10^{11}$ years ago, which for various reasons seemed unacceptable." The "various reasons" in the quote is almost certainly a reference to the fact that the estimated timespan of dynamic models was not larger than the ages of stars as estimated from astrophysics. However, it is possible that Einstein's difficulty also concerns the very idea of a "beginning in time" for the universe.

In the second part of the manuscript, Einstein suggests an alternative solution to the field equations that is also compatible with Hubble's observations – namely, an expanding universe in which the density of matter does not change over time: "In what follows, I wish to draw attention to a solution to equation (1) that can account for Hubbel's [sic] facts, and in which the density is constant over time."

Assuming a metric of flat space expanding exponentially,[7] Einstein derives two simultaneous equations from the field equations, eliminating the cosmological constant to solve for the matter density:

" Equations (1) yield:

$$\frac{-3(9)}{4}\alpha^2 + \lambda c^2 = 0$$

$$\frac{3}{4}\alpha^2 - \lambda c^2 = \kappa\rho c^2$$

*or*

$$\alpha^2 = \frac{\kappa c^2}{3}\rho \qquad .... \qquad (4)"$$

From his equation (4), Einstein concludes that the density of matter $\rho$ remains constant and is related to the expansion factor $\alpha$: "The density is therefore constant and determines the expansion apart from its sign." This would be a stunning result, but it should be noted that equation (4) is incorrect, and arose from a numerical error in the derivation of the coefficient of $\alpha^2$ in the first of the simultaneous equations. Careful study of the manuscript shows that Einstein amended this coefficient from +9/4 to -3/4 (see figure 2), an amendment that leads to the null solution $\rho = 0$ instead of equation (4).

[6] The blank space representing theoreticians other than Tolman is puzzling as Einstein was unquestionably aware of the dynamic models of Friedman and Lemaître.
[7] It is easily shown that assumptions of homogeneity and isotropy imply this metric for a steady-state model.

In the final paragraph of the manuscript, Einstein proposes a physical mechanism to allow the density of matter to remain constant in an expanding universe, namely the continuous formation of matter from empty space: "If one considers a physically bounded volume, particles of matter will be continually leaving it. For the density to remain constant, new particles of matter must be continually formed within that volume from space." This proposal anticipates the later 'creation field' of Fred Hoyle in some ways (see section 4). However, Einstein has not introduced a term representing the creation process into the field equations (unlike Hoyle). Instead, Einstein proposes that the cosmological constant assigns an energy to empty space that can be associated with the creation of matter: "The conservation law is preserved in that, by setting the λ-term, space itself is not empty of energy; its validity is well-known to be guaranteed by equations (1)." Thus, Einstein associates the continuous formation of matter from empty space with the cosmological constant. In reality, the lack of a specific term representing matter creation leads to a universe without matter in this model. It appears that Einstein recognized this problem on revision of the manuscript and set the model aside without pursuing the matter further.

## 4. On steady-state models of the universe

The concept of a continuous creation of matter arose many times in 20th century cosmology. In 1918, the American physicist William MacMillan proposed a continuous creation of matter from radiation in order to avoid a gradual 'running down' of the universe due to the conversion of matter into energy in stellar processes (MacMillan 1918, 1925). The proposal was welcomed by Robert Millikan, who suggested that the process might be the origin of cosmic rays (Millikan 1928). The idea of a continuous creation of matter from radiation was also briefly considered by Richard Tolman as a means of introducing matter into the empty de Sitter universe, although he found the idea improbable (Tolman 1929).

Other physicists considered the possibility of a continuous creation of matter from empty space. In 1928, James Jeans speculated that matter was continuously created in the centre of the spiral nebulae (Jeans 1928) and similar ideas of continuous creation were explored by Svante Arrhenius and Walther Nernst (Arrhenius 1908; Nernst 1928).[8]

Following the discovery of the systematic recession of the spiral nebulae, Richard Tolman suggested that a continuous annihilation of matter into radiation might be responsible for an expansion of space (Tolman 1930a). While Eddington took the view that this process would retard expansion (Eddington 1930), it is possible that Tolman's paper provided the inspiration for Einstein's steady-state model. As pointed out by Harry Nussbaumer, Einstein had many conversations with Tolman at the relevant time and Einstein's steady-state manuscript bears some mathematical similarities to Tolman's model - if not matter annihilation, why not matter creation? (Nussbaumer 2014a).

The concept of an expanding universe that remains in a steady state due to a continuous creation of matter from empty space is most strongly associated with the Cambridge physicists Fred Hoyle, Hermann Bondi and Thomas Gold (Hoyle 1948; Bondi and Gold 1948). In the late 1940s, these physicists became concerned with well-known problems associated with evolving models of the cosmos. In particular, they noted that the evolving models predicted a timespan for expansion that was problematic and disliked Lemaître's hypothesis of a universe with a fireworks beginning (Lemaître 1931a, 1931b, 1931c). Another concern was philosophical in nature; if the universe was truly different in the past, was it not inconsistent to assume that today's laws of physics applied? In order to circumvent these problems, the trio explored the idea of an expanding universe that does not evolve over time, i.e., an expanding cosmos in which the mean density of matter is maintained constant by a continuous creation of matter from the vacuum (Hoyle 1948; Bondi and Gold 1948).

In the case of Bondi and Gold, the proposal of a steady-state model took as starting point the 'perfect cosmological principle', a philosophical principle that stated that the universe should appear essentially the same to all observers in all locations at all times. This principle demanded a continuous creation of matter in order to maintain a constant density of matter in the expanding universe. The resulting model bore some similarities to Einstein's steady-state model, but it is difficult to compare the theories directly as the Bondi-Gold theory was not formulated in the framework of general relativity. On the other hand, Fred Hoyle constructed a steady-state model of the cosmos by means of a daring modification of the Einstein field equations (Hoyle 1948; Mitton 2005 pp 118-119). Replacing Einstein's cosmological constant with a new 'creation-field' term $C_{ik}$ to represent a continuous formation of matter from the vacuum, Hoyle obtained the equation

$$\left(R_{ik} - \frac{1}{2} g_{ik} R\right) - C_{ik} = \kappa T_{ik}$$

[8] A review of steady-state cosmologies in the early 20th century can be found in (Kragh 1996) pp 143-162 .

Hoyle's creation-field term allowed for an unchanging universe but was of importance only on the largest scales, in the same manner as the cosmological constant. In this model, the expansion of space was driven by the creation of matter, and the perfect cosmological principle emerged as consequence rather than starting assumption. A more sophisticated formulation of the model, based on the principle of least action, was proposed in later years (Hoyle and Narlikar 1962).

As is well known, a significant debate was waged between steady-state and evolving models of the cosmos during the 1950s and 1960s (Kragh 1996 pp 252-268; Kragh 2007 pp 187-190; Mitton 2005 pp 167-196). Eventually, the steady-state universe was effectively ruled out by observation, not least by the study of the distribution of the galaxies at different epochs and by the discovery of the cosmic microwave background (Kragh 1996 pp 318-380; Kragh 2007 pp 201-206; Narlikar 1988 pp 218-219). There is no evidence that any of the steady-state theorists were aware of Einstein's attempt; indeed, it is likely that they would have been greatly intrigued to learn that Einstein had once considered a steady-state model.

## 5. On Einstein's philosophy of cosmology

It should come as no great surprise that, when confronted with empirical evidence for an expanding universe, Einstein considered a steady-state or 'stationary' model of the expanding cosmos. Such a model fits well with his lack of interest in non-static solutions to the field equations in 1917, and his hostility to the dynamic models of Friedman and Lemaître when they were first proposed (see section 2). Indeed, a model of the expanding cosmos in which the mean density of matter remains unchanged over time seems a natural successor to Einstein's static model of 1917 from a philosophical point of view.

However, Einstein's attempt at a steady-state model led to a null solution, and it appears that he abandoned the idea rather than pursue it further. (One possibility would have been to introduce a matter-creation term to the field equations in the manner of Hoyle; another to consider a fluid of negative pressure (McCrea 1951)). Instead, Einstein turned to expanding models of varying matter density that could be described 'naturally' by the field equations, i.e., without the use of the cosmological constant term (Einstein 1931b; Einstein and de Sitter 1932). It therefore seems very likely that Einstein abandoned steady-state cosmology on the grounds that it was more contrived than evolutionary models of the cosmos.

Taken together, Einstein's abandonment of steady-state cosmology, his removal of the cosmological constant term in the Friedman-Einstein model (Einstein 1931b), and the removal of spatial curvature in the Einstein-de Sitter model (Einstein and de Sitter 1932), suggest a simple, pragmatic approach to cosmology. Where theorists such as Friedman, Heckmann and Robertson considered all possible universes (see section 2), Einstein sought the simplest model of the universe that could account for observation. It is worth asking whether this practical 'Occam's razor' approach was in fact characteristic of Einstein's cosmology all along, as considered below.

### *5.1 Einstein's journey from the static to the evolving universe*

Einstein's journey from a static, bounded cosmology to the evolving universe is traditionally characterized as that of a reluctant convert; a conservative Einstein, hidebound by philosophical prejudice until overwhelmed by irrefutable evidence (Kragh 1996 p26; Giulini and Straumann 2006; Nussbaumer and Bieri 2009 pp 92; Nussbaumer 2014b; Smeenk 2014). We suggest that Einstein's steady-state manuscript provides a useful clue that this narrative may be somewhat inaccurate.

Considering first Einstein's cosmic model of 1917, it is often asserted that the cosmological constant was introduced to the field equations in order to predict a static rather than a contracting universe. In fact, it is more accurate to say that the purpose of the cosmological constant was to allow the prediction of a finite density of matter in a universe that was assumed *a priori* to be static. No evidence for a dynamic universe was known at the time, and the notion of an expanding or contracting universe would have seemed very far-fetched. (Indeed, Einstein refers to the model as "making possible a quasi-static distribution of matter, as required by the fact of the small velocities of the stars" (Einstein 1917b)).When Friedman explored time-varying solutions of the field equations as a hypothetical possibility in 1922, Einstein was one of the few who paid attention; however, he found non-static solutions 'suspicious' due to a lack of supporting evidence (see section 2). In 1927, Lemaître's expanding model of the universe was inspired by observations at the cutting edge of astronomical research; Einstein's rejection of this model can probably be attributed to a lack of familiarity with advances in astronomy. Lemaître certainly thought so, commenting later that Einstein did not seem to be aware of recent astronomical measurements (Lemaître 1958).

With the publication of astronomical observations suggestive of an expanding cosmos in 1929, Einstein lost little time in abandoning the static universe. It seems that he had no difficulty changing his viewpoint once such a change was warranted by the evidence. One is reminded of a famous comment attributed to John Maynard Keynes: "When the facts change, I change my mind - what do you do, Sir?" It now seems that at this point, Einstein's first guess was an expanding universe that remains essentially unchanged over time - the obvious next step after his static model. However, when this attempt led to an empty universe, Einstein turned to evolving models instead. Noting that expanding models did not necessarily require a cosmological constant, he removed this term (Einstein 1931b). When he realised that spatial curvature was also no longer a given in dynamic cosmologies, this parameter was removed in turn (Einstein and de Sitter 1932). This sequence of ever simpler models suggests an approach to cosmology that was not conservative but pragmatic - a minimalist, empirical approach to the study of the universe. Tellingly, Einstein did not propose any major cosmic models beyond this point; as he explained later, he saw little point in speculating further in the absence of empirical data on cosmological parameters such as spatial curvature and the density of matter (Einstein 1945 pp 133-134).

We note that this approach to cosmology is very typical of Einstein's general approach to physics, at least in his younger years. Sometimes described as positivist, Einstein's approach is more accurately described as a philosophy of *logical empiricism* – he embraced the central importance of observations in the testing of a theoretical hypothesis, at least in a holistic sense, but also assigned great importance to the construction of consistent theories from analytic principles of logic (Frank 1948, pp 259-263; Frank 1949 pp 271-286; Reichenbach 1949 pp 309-311; Einstein 1949 pp 680-681). This is a very different approach to that of Compte or Mach, who suggested that the fundamental laws of physics should only contain concepts that could be defined by direct observations, or at least be connected to observation by a short chain of thought. It is also different to that of empiricists such as Moritz Schlick or Rudolf Carnap because it contained both positivist and metaphysical elements.[9] An insight into Einstein's philosophy of science in these years can be found in his 1933 Herbert Spencer Lecture at Oxford: "Experience remains, of course, the sole criterion of the physical utility of a mathematical construction. But the creative principle resides in mathematics" (Einstein 1933b; Einstein 1934 p36).

[9] See (Howard 2014) for an overview of this point.

### *5.2 On the cosmological constant and dark energy*

Until recently, it was universally assumed that, with the emergence of the first empirical evidence for an expanding universe, Einstein immediately abandoned the cosmological constant along with the static universe (North 1965 p132; Kragh 1996 p34; Straumann 2002; Nussbaumer and Bieri 2009 p147; Nussbaumer 2014b). Certainly, Einstein made it clear on several occasions that he disliked the term, at least from the perspective of the general theory of relativity. (For example, in 1919 he described the term as "gravely detrimental to the formal beauty of the theory" (Einstein 1919)). However, Einstein's steady-state manuscript demonstrates that he retained the cosmological constant in at least one cosmic model he attempted *after* Hubble's observations, albeit for a new purpose. It appears that when presented with evidence for a cosmic expansion, Einstein's attraction to an unchanging universe at first outweighed his dislike of the cosmological constant, just as it did in 1917.

It will not have escaped the reader's attention that Einstein's association of the cosmological constant with an energy of space in his steady-state model is not unlike today's hypothesis of dark energy, at least from a philosophical standpoint. Where Einstein attempted to associate a continuous creation of matter with the cosmological constant, today we assume an energy for an accelerated expansion.[10] More generally, it has often been noted that the cosmological constant term of 1917 anticipates the notion of dark energy in some ways. It is less well-known that Einstein also considered – and dismissed - the possibility of a time-varying energy of space, a concept not unlike the modern hypothesis of quintessence. Within a few months of the publication of Einstein's static model of 1917, Erwin Schrödinger suggested that the cosmological term could be placed on the right hand side of the field equations (a negative energy density term in the matter-energy tensor) and that the term could be time-varying (Schrödinger 1918). Einstein's response was that, if constant, placing the term in the matter-energy tensor was equivalent to his original formulation. If not constant, the term would necessitate undesirable speculation on the nature of its variation over time: "The course taken by Herr Schrödinger does not appear passable to me because it leads too deeply into the thicket of hypotheses" (Einstein 1918c). Once again, this attitude indicates a strong dislike of complicated solutions unless necessitated by observation.[11]

[10] See (Peebles and Ratra 2003) for a review of dark energy.

We note that a great deal has been written over the years about Einstein's evolving view of the cosmological constant. For example, the well-known Russian physicist George Gamow stated that Einstein once declared the term "my greatest blunder" (Gamow 1956; Gamow 1970 p44), while others have cast doubt on this statement (Straumann 2002; Livio 2013 pp 233-241). We will not enter this debate here, but simply note that Einstein soon dispensed with the term in his non-static cosmology. His considered view is probably best summed up in a footnote to his 1945 review of cosmology: "If Hubble's expansion had been discovered at the time of the creation of the general theory of relativity, the cosmologic member would never have been introduced. It seems now so much less justified to introduce such a member into the field equations, since its introduction loses its sole original justification – that of leading to a natural solution of the cosmologic problem" (Einstein 1945 p130). This stance should be contrasted with Einstein's attitude to spatial curvature. While the Einstein-de Sitter model was based on the fact that the presence of matter in a dynamic universe does not automatically imply spatial curvature, the authors were careful not to rule it out: "It is possible to represent the facts without assuming a curvature of three-dimensional space. The curvature is, however, essentially determinable, and an increase in the precision of the data derived from observations will enable us in the future to fix its sign and to determine its value" (Einstein and de Sitter 1932).

### *5.3 On the question of cosmic origins*

To modern eyes, a striking aspect of Einstein's steady-state manuscript is the lack of reference to the problem of the singularity for the case of evolving models, or to the related question of an origin for the universe. Indeed, the manuscript is the only steady-state model of the expanding universe known to us that is not motivated (at least in part) by a desire to circumvent such problems. While Einstein is clearly conscious of the puzzle of the short timespan of evolving models, there is no reference to the problem of origins (see section 3).

One explanation might be that Einstein's steady-state manuscript almost certainly pre-dated Lemaître's proposal of a 'fireworks beginning' for the universe (Lemaître 1931b, 1931c). However, the issue of cosmic origins for evolving models was recognized before these papers were published (Eddington 1930, 1931; de Sitter 1932). We note instead that

[11] See (Harvey 2012) for a fuller discussion of this episode.

Einstein's silence on the question is very typical of his cosmology – there is no reference to the problem in either of his evolving models (Einstein 1931b; Einstein and de Sitter 1932) or in a contemporaneous review of relativistic cosmology (Einstein 1933a). In later years, Einstein made it clear that this silence did not stem from a philosophical difficulty with the notion of a physical origin for the cosmos, but from doubts concerning the validity of relativistic models at early epochs: "For large densities of field and of matter, the field equations and even the field variables which enter into them will have no real significance. One may not therefore assume the validity of the equations for very high density of field and of matter" (Einstein 1945 pp 132-133).

### *5.4 On Einstein's philosophy of relativity*

We note in passing that Einstein's steady-state manuscript does not contain any considerations of philosophical issues associated with the theory of relativity, as opposed to cosmology. Reading the opening section of the work, the professional philosopher may be somewhat disappointed by the lack of reference to problems such as the use of idealised clocks and rulers in relativity,[12] or the question of the geometrization of gravity.[13] This silence is once again very typical of Einstein's cosmology; such issues are not discussed in any of Einstein's static or dynamic models of the cosmos, although he did consider them elsewhere (Einstein 1948). This suggests once more that Einstein's approach to cosmology was essentially pragmatic; general relativity was a useful tool to describe the universe, but by no means the ultimate answer. As we have argued elsewhere (O'Raifeartaigh and McCann 2014), it is likely that Einstein's search for a unified field theory in these years made him very conscious of the limitations of relativistic models of the cosmos.

### *5.5 On paradigm shifts in cosmology*

We note finally that Einstein's steady-state manuscript does not support a view that his acceptance of the evolving universe occurred as an abrupt change to a new worldview. As described above, the model appears as an intermediate step in a long journey from the static universe to an expanding, evolving cosmology. Indeed, the manuscript provides a new piece of evidence that today's 'big bang' cosmology did not emerge as an abrupt 'paradigm shift' in the manner envisioned by Thomas Kuhn (Kuhn 1962), but rather as a slow dawning in both theory and observation within a single paradigm, the relativistic universe.

[12] See (Brown 2014) for a review.
[13] A longstanding question was whether the spacetime metric of relativity was a mathematical tool to describe gravity, or whether gravity 'was' geometry (Lehmkuhl 2014).

It is unfortunate that Einstein's cosmology papers of the 1930s are not better known, as the pragmatic, empirical approach we have discussed above is very different to Einstein's work on unified field theory in these years (Einstein and Mayer 1930, 1931, 1932). Indeed, we find the cosmology papers quite reminiscent of the young Einstein's approach to emerging phenomena (Einstein 1905a, 1905b, 1905c). One wonders whether the familiar narrative that Einstein became more and more attached to a formal mathematical approach to physics in his later years is entirely accurate. Could it be that Einstein's philosophical approach to science did not truly change, but that the intense level of mathematical abstraction one associates with Einstein's later work was simply a facet of the great technical challenge posed by unified field theory?

**6. Conclusions**

Einstein's attempt at a steady-state model was abandoned before publication but it offers many insights into his philosophy of cosmology. His hypothesis of a universe of expanding radius and constant matter density is very different to his static model of 1917 or his evolving models of 1931 and 1932, and anticipates in some ways the well-known steady-state cosmology of Hoyle, Bondi and Gold. The model was almost certainly written in early 1931, when Einstein first learnt of observational evidence for a cosmic expansion, but was quickly abandoned when it led to a null solution. The steady-state manuscript is nevertheless of interest because it offers new evidence that Einstein's philosophical journey from a static, bounded cosmology to the dynamic, evolving universe was that of a pragmatic empiricist, rather than a reluctant conservative.

We note finally that Einstein's steady-state model finds an echo in today's theories of cosmic inflation. In particular, the de Sitter metric of flat, exponentially expanding space used in inflationary models[14] recalls the steady-state models of Einstein and Hoyle. Indeed, many scholars have noted that inflationary models are effectively steady-state cosmologies over an extremely limited timespan (Hoyle 1994 p271; Narlikar 1988 pp 223-225; Narlikar 2005; Barrow 2005). Further, it has been suggested (Vilenkin 1983; Linde 1986a,1986b) that the inflationary process inevitably creates the conditions for further inflation in a never-ending cycle. This concept of 'eternal inflation' raises the possibility that the observed, evolving universe is a local anomaly in a global ensemble that is in a steady state (Barrow 2005), a scenario that is not dissimilar to Hoyle's later proposal of a steady-state universe permeated with local 'little bangs' (Hoyle and Narlikar 1966; Hoyle, Burbidge and Narlikar 1993; Narlikar 2005). Thus it can be concluded that, like the cosmological constant, the concept of the steady state universe is proving hard to banish from modern cosmology.

[14] See (Liddle 1999) for a review of inflationary cosmology.

## Acknowledgements

The author would like to thank the Hebrew University of Jerusalem for permission to display the excerpts shown in figures 1 and 2. He also thanks the Dublin Institute of Advanced Studies for access to the Collected Papers of Albert Einstein (Princeton University Press).

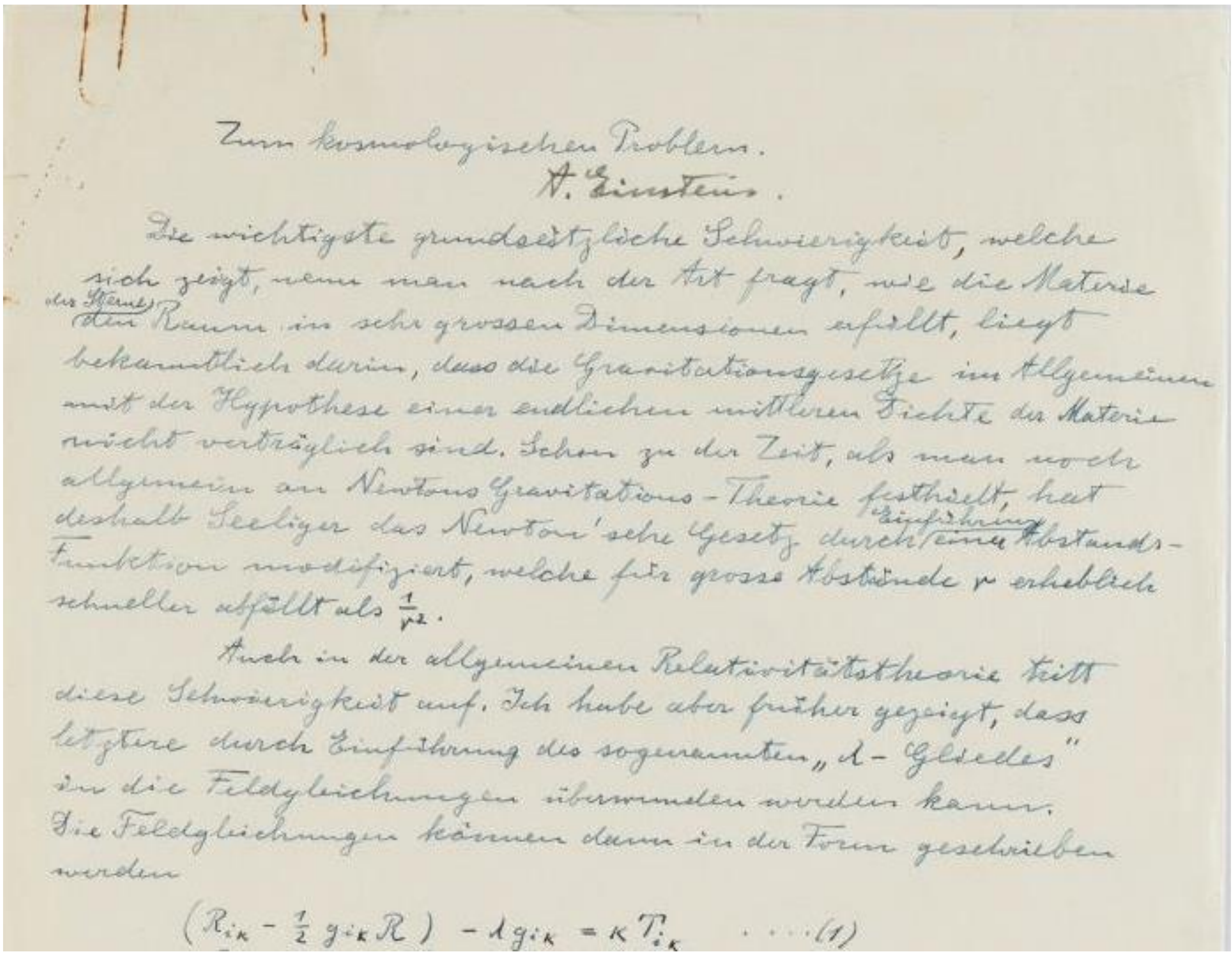

Zum kosmologischen Problem.

A. Einstein.

Die wichtigste grundsätzliche Schwierigkeit, welche sich zeigt, wenn man nach der Art fragt, wie die Materie der Sterne Raum in sehr grossen Dimensionen erfüllt, liegt bekanntlich darin, dass die Gravitationsgesetze im Allgemeinen mit der Hypothese einer endlichen mittleren Dichte der Materie nicht verträglich sind. Schon zu der Zeit, als man noch allgemein an Newtons Gravitations-Theorie festhielt, hat deshalb Seeliger das Newton'sche Gesetz durch Einführung Abstands-Funktion modifiziert, welche für grosse Abstände $r$ erheblich schneller abfällt als $\frac{1}{r^2}$.

Auch in der allgemeinen Relativitätstheorie tritt diese Schwierigkeit auf. Ich habe aber früher gezeigt, dass letztere durch Einführung des sogenannten „$\lambda$-Gliedes" in die Feldgleichungen überwunden werden kann. Die Feldgleichungen können dann in der Form geschrieben werden

$$\left(R_{ik} - \frac{1}{2} g_{ik} R\right) - \lambda g_{ik} = \kappa T_{ik} \quad \ldots (1)$$

**Figure 1.** An excerpt from the first page of Einstein's steady-state manuscript (Einstein 1931a). Reproduced by kind permission of the Hebrew University of Jerusalem.

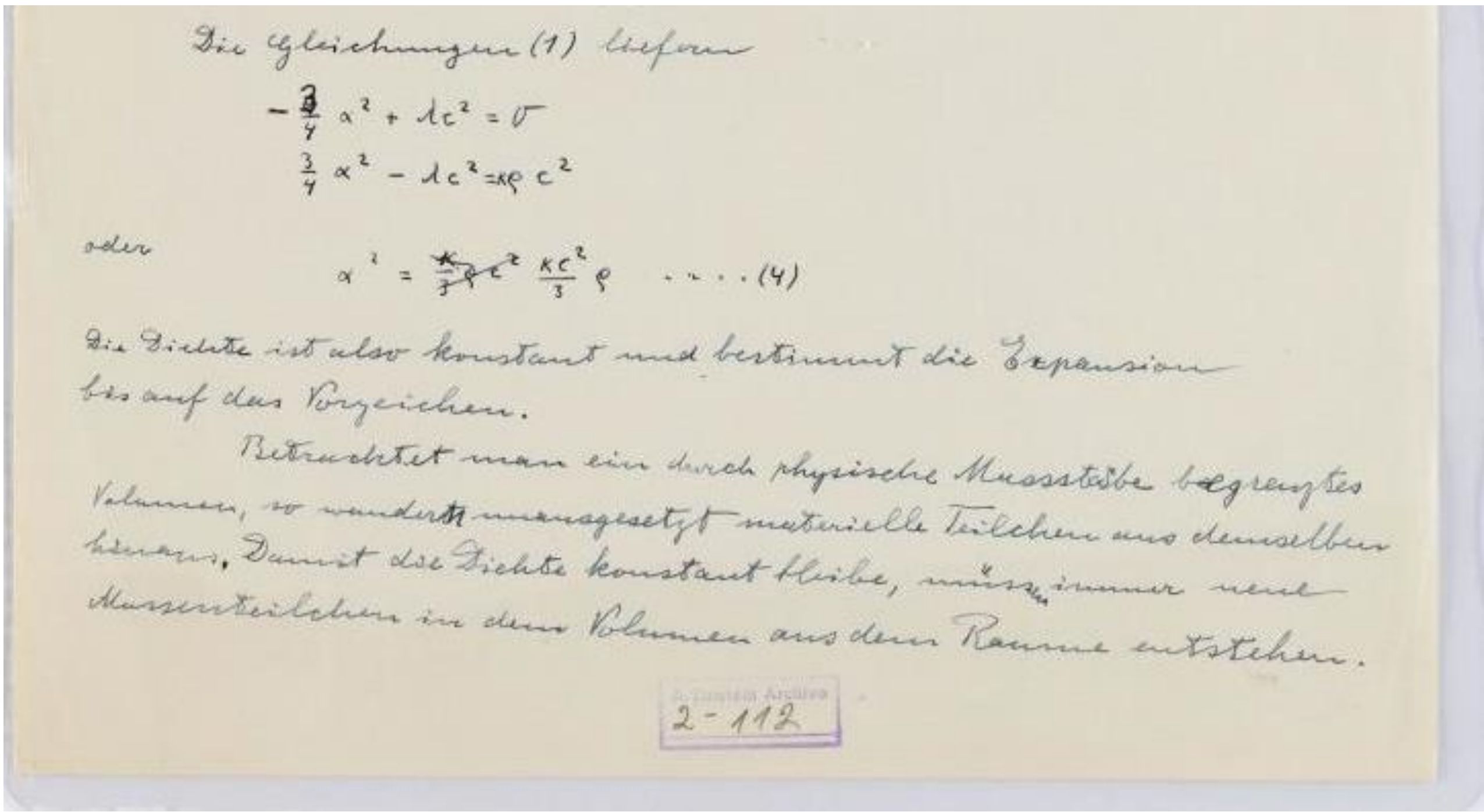
Die Gleichungen (1) liefern

$$-\frac{3}{4}\alpha^2 + \lambda c^2 = 0$$

$$\frac{3}{4}\alpha^2 - \lambda c^2 = \kappa\rho c^2$$

oder

$$\alpha^2 = \frac{\kappa c^2}{3}\rho \quad \ldots\ldots (4)$$

Die Dichte ist also konstant und bestimmt die Expansion bis auf das Vorzeichen.

Betrachtet man ein durch physische Massstäbe begrenztes Volumen, so wandern fortgesetzt materielle Teilchen aus demselben heraus. Damit die Dichte konstant bleibe, müssen immer neue Masseteilchen in dem Volumen aus dem Raume entstehen.

**Figure 2.** An excerpt from the last page of Einstein's steady-state manuscript (Einstein 1931a), reproduced by kind permission of the Hebrew University of Jerusalem. Equation (4) implies a direct relation between the expansion coefficient α and the mean density of matter ρ. However, the coefficient of $\alpha^2$ in the first of the simultaneous equations was amended from 9/4 to -3/4 on revision, a correction that gives the null result $\rho = 0$ instead of equation (4).